\documentclass[reprint,showpacs,preprintnumbers,amsmath,aps,pre]{revtex4-1}
\usepackage{graphicx}
\begin{document}

\title{Coefficient of performance at maximum figure of merit and its bounds for low-dissipation Carnot-like refrigerators}
\author{Yang Wang}
\affiliation{Department of Physics, Beijing Normal University, Beijing 100875, China}
\author{Mingxing Li}
\affiliation{Department of Physics, Beijing Normal University, Beijing 100875, China}
\author{Z. C. Tu}\email{Corresponding author. Email: tuzc@bnu.edu.cn}
\affiliation{Department of Physics, Beijing Normal University, Beijing 100875, China}
\author{A. Calvo Hern\'{a}ndez}
\affiliation{Departamento de F\'{\i}sica Aplicada, and Instituto Universitario de F\'{\i}sica Fundamental y Matem\'{a}ticas (IUFFyM),
Universidad de Salamanca, E-37008 Salamanca, Spain}
\author{J. M. M. Roco}
\affiliation{Departamento de F\'{\i}sica Aplicada, and Instituto Universitario de F\'{\i}sica Fundamental y Matem\'{a}ticas (IUFFyM),
Universidad de Salamanca, E-37008 Salamanca, Spain}

\date{\today}

\begin{abstract}
The figure of merit for refrigerators performing finite-time Carnot-like cycles between two reservoirs at temperature $T_h$ and $T_c$ ($<T_h$) is optimized. It is found that the coefficient of performance at maximum figure of merit is bounded between $0$ and $(\sqrt{9+8\varepsilon_c}-3)/2$ for the low-dissipation refrigerators, where $\varepsilon_c =T_c/(T_h-T_c)$ is the Carnot coefficient of performance for reversible refrigerators. These bounds can be reached for extremely asymmetric low-dissipation cases when the ratio between the dissipation constants of the processes in contact with the cold and hot reservoirs approaches to zero or infinity, respectively. The observed coefficients of performance for real refrigerators are located in the region between the lower and upper bounds, which is in good agreement with our theoretical estimation.
\end{abstract}
\maketitle

\emph{Introduction}.-- The issue of efficiency at maximum power output has attracted much attention since the seminal achievements made by Yvon~\cite{Yvon55}, Novikov~\cite{Novikov}, Chambadal~\cite{Chambadal}, Curzon and Ahlborn~\cite{Curzon1975}, which gives rise to finite-time thermodynamics, a new branch of non-equilibrium thermodynamics, and opens open new avenues to the perspective of establishing more realistic theoretical bounds for real heat engines as well as refrigerators~\cite{Chen1989,ChenJC94,Bejan96,ChenL99}.

Previous reported works on this subject show that different model systems exhibit various kinds of behaviors at large relative temperature difference between two thermal reservoirs at temperatures $T_h$ and $T_c~(<T_h)$, in spite that they show certain universal behavior at small relative temperature difference~\cite{vdbrk2005,dcisbj2007,Izumida,Schmiedl2008,Tu2008,Esposito2009a,Esposito2009,wangx11,Apertet12} leading to recent discussions on the bounds of efficiency at maximum power output for Carnot-like heat engines~\cite{EspositoPRE10,Esposito2010,Velasco10,GaveauPRL10,WangTu2011,wangtu2012}. In particular, Esposito \textit{et al.} investigated low-dissipation Carnot-like engines by assuming that the irreversible entropy production in each isothermal process is inversely proportional to the time required for completing that process~\cite{Esposito2010}. Furthermore, they obtained that the efficiency at maximum power output for low-dissipation engines is bounded between $\eta_-\equiv \eta_C/2$ and $\eta_+\equiv \eta_C/(2-\eta_C)$~\cite{Esposito2010}, where $\eta_C =1-T_c/T_h$ is the Carnot efficiency of reversible heat engines. Besides, Ma~\cite{SKma1985} proposed the per-unit-time efficiency to be another criterion, which can be viewed as a compromise between the efficiency and the speed of the whole cycle. Two of present authors and their coworkers~\cite{ACHernandez1998} proved that the efficiency of endoreversible heat engines performing at maximum per-unit-time efficiency is bounded between $\eta_C/2$ and $1-\sqrt{1-\eta_C}$.

However, it is relatively difficult to define an optimal criterion and obtain its corresponding coefficient of performance (COP) for refrigerators~\cite{YanChen1990,Jincan1998, Jizhouhe,dcisbj2006,Mahler, RocoPRE12,Velasco1997,Lingenchen1995,CSWK1997,Turkey} in the way as we address the issue of efficiency at maximum power for heat engines provided that minimum power input is not an appropriate figure of merit in Carnot-like refrigerators. Velasco \textit{et al.}~\cite{Velasco1997} adopted the per-unit-time COP as an target function and proved $\varepsilon_{CA}\equiv\sqrt{\varepsilon_c+1}-1$ to be the upper bound of COP for endoreversible refrigerators operating at the maximum per-unit-time COP, being $\varepsilon_c =T_c/(T_h-T_c)$ the Carnot COP for reversible refrigerators. Allahverdyan \emph{et al.} \cite{Mahler} investigated a quantum model which consists of two $n$-level systems interacting via a pulsed external field and took $\varepsilon Q_c$ as the target function, where $\varepsilon$ and $Q_c$ are the COP of refrigerators and the heat absorbed from the cold reservoir, respectively. They also proved that the COP of this model is bounded between $\varepsilon_{CA}$ and $\varepsilon_c$ at the small relative temperature difference. Chen and Yan \cite{YanChen1990} suggested to take $\chi=\varepsilon Q_c/t_{\mathrm{cycle}}$ as the target function, where $t_{\mathrm{cycle}}$ is the time for completing the whole Carnot-like cycle. Recently, de Tom\'{a}s and two of present authors \cite{RocoPRE12} optimized $\chi$ for symmetric low-dissipation refrigerators and derived the COP at maximum $\chi$ to be $\varepsilon_{CA}=\sqrt{1+\varepsilon_c}-1$.
The above results give rise to two straightforward questions: (i) What target function could be appropriate as the figure of merit for refrigerators? (ii) Can we derive the bounds of COP at maximum figure of merit for general low-dissipation refrigerators as a counterpart to the bounds of efficiency at maximum power output for heat engines?
We will address these problems in this work. We select $\chi=\varepsilon Q_c/t_{\mathrm{cycle}}$ as the figure of merit and derive that the COP at maximum figure of merit is bounded between $0$ and $(\sqrt{9+8\varepsilon_c}-3)/2$ for low-dissipation refrigerators. Our theoretical prediction is in good agreement with the observed data from real refrigerators, which suggests $\chi=\varepsilon Q_c/t_{\mathrm{cycle}}$ is appropriate as the figure of merit for refrigerators.

\emph{Model}.--The refrigerator that we consider performs a Carnot-like cycle consisting of two isothermal processes and two adiabatic steps as follows. It must be noted that the word ``isothermal" in this work also merely indicates that the working fluid is in contact with a reservoir at constant temperature. Here we do not introduce the effective temperature of working fluid because the effective temperature might not be well-defined in many cases~\cite{wangtu2012}.

(S1)~Isothermal expansion. The working substance is in contact with a cold reservoir at temperature $T_c$ and the constraint on the system is loosened according to the external controlled parameter $\lambda_c(\tau)$ during the time interval $0<\tau<t_c$, where $\tau$ is a time variable. It is in the sense of loosening the constraint that this step is called an expansion process. A certain amount of heat $Q_c$ is absorbed from the cold reservoir. Then the variation of entropy in this process can be expressed as
\begin{equation}
\Delta S_c={Q_c}/{T_c} +\Delta S_c^{ir},
\label{Eq-deltaS1}
\end{equation}
where $\Delta S_c^{ir}\ge 0$ is the irreversible entropy production. We adopt the convention that the heat absorbed by the refrigerator is positive, so $\Delta S_c^{ir} \le\Delta S_c$.

(S2)~Adiabatic compression. This step is idealized as the working substance suddenly decouples from the cold reservoir and then comes into contact with the hot reservoir instantaneously at time $t_{c}$. During this transition, the controlled parameter is switched from $\lambda_c(t_c)$ to $\lambda_h(t_h)$ [$>\lambda_c(t_c)$], that is, the constraint on the system is enhanced. It is in the sense of enhancing the constraint that this step is called a compression process.
There is no heat exchange in this transition, i.e. $Q_2=0$. The distribution function of molecules of working substance is unchanged. Thus there is no entropy production in this transition, i.e. $\Delta S_2=0$.

(S3)~Isothermal compression. The working substance is in contact with a hot reservoir at temperature $T_h$ and the constraint on the system is further enhanced according to the external controlled parameter $\lambda_h(\tau)$ during the time interval $t_c<\tau<t_{c}+t_h$. A certain amount of heat $Q_h$ is released to the hot reservoir $T_h$. Thus the total variation of entropy in this process is
\begin{equation}
\Delta S_h = -{Q_h}/{T_h}+\Delta S_h^{ir},
\label{Eq-deltaS3}
\end{equation}
where $\Delta S_h^{ir}\ge 0$ is the irreversible entropy production.

(S4)~Adiabatic expansion.
Similar to the adiabatic compression process, the working substance suddenly decouples from the hot reservoir and then comes into contact with the cold reservoir instantaneously at time $t_{c}+t_{h}$. During this transition, the controlled parameter is switched from $\lambda_h (t_c+t_h)$ to $\lambda_c (0)$ [$<\lambda_h (t_c+t_h)$], that is, the constraint on the system is loosened.
In this transition, both the heat exchange and the entropy production are vanishing, i.e. $Q_4=0$ and $\Delta S_4=0$.

\emph{Optimizing the figure of merit}.--Having undergone a whole cycle, the system recovers its initial state. Thus the change of entropy is vanishing for the whole cycle, from which we can easily derive that the variations of entropy in two ``isothermal" processes satisfy $\Delta S_c =-\Delta S_h \equiv \Delta S>0$. Similarly, the total energy also remains unchanged for the whole cycle, thus the work input in the cycle can be expressed as $W=Q_h-Q_c$, and then the COP of refrigerators is reduced to
\begin{equation}\varepsilon=Q_c/(Q_h-Q_c).\label{Eq-COPdef}\end{equation}
Considering Eqs.~\eqref{Eq-deltaS1}--\eqref{Eq-COPdef} and $t_{\mathrm{cycle}}=t_c+t_h$, the figure of merit $\chi\equiv \varepsilon Q_c/t_{cycle}$ is transformed into
\begin{equation}\label{targerchir}
    \chi=\frac{T_c^2(\Delta S-\Delta S_c^{ir})^2}{[(T_h-T_c)\Delta S +T_c \Delta S_c^{ir}+T_h\Delta S_h^{ir}](t_h+t_c)}.
\end{equation}
The variation of entropy $\Delta S$ is a state variable only depending on the initial and final states of the isothermal processes while $\Delta S^{ir}_c$ and $\Delta S^{ir}_h$ are process variables relying on the detailed protocols $\lambda(\tau)$. In addition, $\Delta S^{ir}_c<\Delta S$ according to Eq.~\eqref{Eq-deltaS1}. Thus Eq.~\eqref{targerchir} implies that the maximum of the figure of merit corresponds to minimizing irreversible entropy production $\Delta S_c^{ir}$ and $\Delta S_h^{ir}$ with respect to the protocols for given time intervals $t_c$ and $t_h$, which is equivalent to that obtained for Carnot-like heat engines working at maximum power output.

To continue our analysis, we denote the minimum irreversible entropy production with the optimized protocols as $\min\{\Delta S_c^{ir}\}\equiv L_c(t_c)$ and $\min\{\Delta S_h^{ir}\}\equiv L_h(t_h)$. Intuitively, $L_c(t_c)$ and $L_h(t_h)$ are the monotonous decreasing functions of $t_c$ and $t_h$, respectively, because the larger time for completing the isothermal steps, the closer these steps are to quasistatic processes so that the irreversible entropy production $\Delta S_c^{ir}$ and $\Delta S_h^{ir}$ become much smaller. In particular, $\Delta S_c^{ir}$ and $\Delta S_h^{ir}$ should vanish in the long-time limit $t_c \rightarrow \infty$ and $t_h \rightarrow \infty$.
For convenience, we can make a variable transformation $x_c=1/t_c$ and $x_h=1/t_h$. If we consider Eqs.~\eqref{Eq-deltaS1} and \eqref{Eq-deltaS3}, the heat $Q_c$ and $Q_h$ can be expressed as
\begin{equation}
Q_c=T_c[\Delta S-L_c(x_c)] \label{eqq_c},\end{equation} and
\begin{equation}Q_h=T_h[\Delta S+L_h(x_h)] \label{eqq_h}.
\end{equation}

Substituting Eqs.~\eqref{eqq_c} and \eqref{eqq_h} into \eqref{Eq-COPdef}, we derive the COP of refrigerators to be
\begin{equation}\label{varepsilon}
\varepsilon =\frac{Q_c}{Q_h-Q_c}=\frac{T_c(\Delta S-L_c)}{(T_h-T_c)\Delta S+T_cL_c+T_hL_h}.
\end{equation}
Considering $t_{\mathrm{cycle}}=t_c+t_h=1/x_c +1/x_h$ and the above equations \eqref{eqq_c}--\eqref{varepsilon}, we optimize the figure of merit $\chi =\varepsilon Q_c/t_{\mathrm{cycle}}$ with respect to $x_h$ and $x_c$ and derive the following two equations:
\begin{eqnarray}
 (Q_h-Q_c)x_h=(2Q_h/Q_c -1)T_cL'_c x_c(x_h+x_c), \label{opt1}\\
 (Q_h-Q_c)x_c=T_h L_h'x_h(x_h+x_c),  \label{opt2}
\end{eqnarray}
where $L_c' \equiv{\mathrm{d} L_c}/{\mathrm{d}x_c}$ and $L_h' \equiv {\mathrm{d} L_h}/{\mathrm{d} x_h}$.

Considering Eqs.~\eqref{eqq_c}--\eqref{varepsilon} and then dividing Eq.~\eqref{opt1} by Eq.~\eqref{opt2}, we can derive that the COP at maximum figure of merit satisfies
\begin{equation}{\varepsilon_\ast}{T_h L_h'x_h^2}=(\varepsilon_\ast+2){T_c L_c'x_c^2}.\label{result1}\end{equation}
Similarly, adding Eq.~\eqref{opt1} and Eq.~\eqref{opt2}, we can derive
\begin{equation}
\frac{1}{\varepsilon_\ast}=\frac{1}{\varepsilon_c}+\frac{1}{N\varepsilon_\ast+(2\varepsilon_c-\varepsilon_\ast)M/(1+\varepsilon_c)}\label{result2}
\end{equation}
with reducing parameters $N=(L_c'x_c+L_h'x_h)/(L_c+L_h)$, $M=L'_cx_c/(L_c+L_h)$ and $\varepsilon_c=T_c/(T_h-T_c)$.

\begin{figure}[!htp]
\includegraphics[width=7cm]{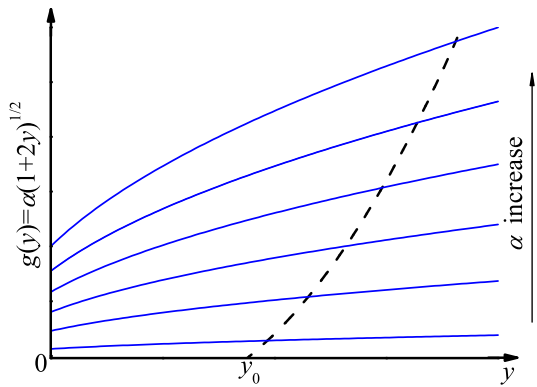}
\caption{(Color online) Schematic diagrams of function. The dashed line is the diagram of $f(y)=2\varepsilon_c y^2-3y-1$ while the solid lines correspond to diagram of $g(y)=\alpha(1+2y)^{1/2}$ with different values of $\alpha$. The points of intersection represent the solutions to Eq.~\eqref{result3} for different values of $\alpha$. $y_0$ is the solution to $f(y)=2\varepsilon_c y^2-3y-1=0$.}\label{graph1}
\end{figure}

\emph{Bounds of COP at maximum figure of merit}.--Now we turn to the low-dissipation refrigerators by assuming that $L_c'=\Sigma_c$ and $L_h'=\Sigma_h$ are two dissipation constants as Esposito \emph{et al.}~\cite{Esposito2010} proposed for low-dissipation heat engines. In this case, $N=1$ and $M=\Sigma_cx_c/(\Sigma_cx_c+\Sigma_hx_h)$. Particularly, $\Sigma_c=\Sigma_h=\Sigma$ for the symmetric low-dissipation cases investigated by de Tom\'{a}s \emph{et al.} \cite{RocoPRE12}, it is not hard for us to recover its COP at maximum maximum figure of merit to be $\varepsilon_{CA}=\sqrt{1+\varepsilon_c}-1$ from Eqs.~\eqref{result1} and \eqref{result2}. However, for the asymmetric low-dissipation cases where $\Sigma_c \neq \Sigma_h$, it is more difficult to obtain a concise analytic expression of $\varepsilon_\ast$ than the symmetric case. But we can still estimate its bounds from Eq.~\eqref{result2}. According to this equation, we have
\begin{equation}\label{varepsilonm}
    \varepsilon_{\ast}=\frac{\varepsilon_c[\sqrt{1+8(1+\varepsilon_c)/M}-3]}{2[(1+\varepsilon_c)/M-1]},
\end{equation}
which is the key equation in the present work. Because $\varepsilon_c > 0$ and $0\le M\le 1$, it is easy to prove that $\varepsilon_\ast$ is a monotonous increasing function of $M$. As a main result, from Eq.~\eqref{varepsilonm} we obtain the wished bounds as:
\begin{equation}\label{bounds}
    0 \le\varepsilon_\ast \le (\sqrt{9+8\varepsilon_c}-3)/2.
\end{equation}

It is noted that $M$ is also constrained by Eq.~\eqref{result1}, which pushes us to further discuss the accessibility of the lower bound $\varepsilon_-\equiv 0$ and the upper bound $\varepsilon_+\equiv (\sqrt{9+8\varepsilon_c}-3)/2$. Eliminating $x_c/x_h$ from Eqs.~\eqref{result1} and \eqref{result2}, we have
\begin{equation}\label{result3}
  2\varepsilon_c y^2-3y-1=\alpha(1+2y)^{1/2},
\end{equation}
where $y=1/\varepsilon_\ast$ and $\alpha=\sqrt{T_h\Sigma_h/T_c\Sigma_c}$.
In Fig.~\ref{graph1}, we schematically plot the function $f(y)=2\varepsilon_c y^2-3y-1$ (dashed line) and $g(y)=\alpha(1+2y)^{1/2}$ for different values of $\alpha$ (solid lines). The points of intersection between the dashed line and solid lines correspond to the solutions to Eq.~\eqref{result3} for different values of $\alpha$.
It follows that the solutions to Eq.~\eqref{result3} increase with the increasing value of $\alpha$. On the other hand,
the values of $\alpha$ can be taken from 0 to $\infty$. Therefore we infer that the solutions to Eq.~\eqref{result3} are between
$y_0=(\sqrt{9+8\varepsilon_c}+3)/4\varepsilon_c$ [solution to Eq.~\eqref{result3} for $\alpha=0$, i.e. ${\Sigma_c}/{\Sigma_h} \rightarrow \infty$] and $\infty$ [solution to Eq.~\eqref{result3} for $\alpha\rightarrow \infty$, i.e. ${\Sigma_c}/{\Sigma_h} \rightarrow 0$]. Noting that $y=1/\varepsilon_\ast$, we arrive at
$0\le \varepsilon_\ast \le 1/y_0 = (\sqrt{9+8\varepsilon_c}-3)/2$ which is exactly
the same as inequality \eqref{bounds}. Simultaneously, we obtain the condition for reaching the lower and upper bounds: $\varepsilon_\ast \rightarrow 0$ when ${\Sigma_c}/{\Sigma_h} \rightarrow 0$ and $\varepsilon_\ast \rightarrow  (\sqrt{9+8\varepsilon_c}-3)/2$ when ${\Sigma_c}/{\Sigma_h} \rightarrow \infty$. That is, the lower and upper bounds of COP at maximum figure of merit can be reached for extremely asymmetric low-dissipation refrigerators.
Although the lower and upper bounds of efficiency at maximum power output can also be reached for extremely asymmetric low-dissipation heat engines \cite{Esposito2010}, the subtle difference is that the lower bound can be reached when ${\Sigma_c}/{\Sigma_h} \rightarrow \infty$ while the upper one can be reached when ${\Sigma_c}/{\Sigma_h} \rightarrow 0$, which is in the inverse situation with respect to the refrigerators. However, this difference is quite reasonable because refrigerators need the input work to pump heat from the cold reservoir while heat engines utilize heat from the hot source to generate work.

\begin{figure}[!htp]
\includegraphics[width=7cm]{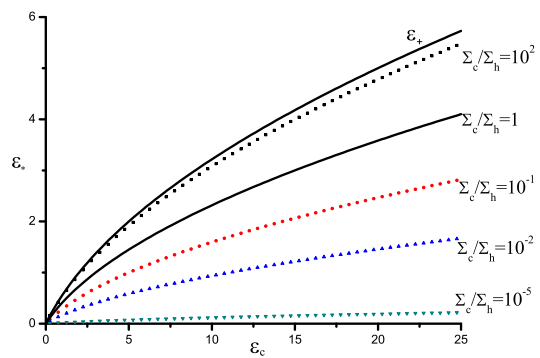}
\caption{(Color online) Numerical solutions to Eq.~\eqref{result3}. The used values of parameter ${\Sigma_c}/{\Sigma_h}$ are marked nearby each curves.}\label{figCOPnum}
\end{figure}

The numerical solutions to Eq.~\eqref{result3} can also be calculated by setting different values of ratio ${\Sigma_c}/{\Sigma_h}$.
The corresponding values of $\varepsilon_\ast =1/ y$ are shown in Fig.~\ref{figCOPnum}, from which we find that the COP at maximum figure of merit indeed reaches the upper bound $\varepsilon_+=(\sqrt{9+8\varepsilon_c}-3)/2$ when the ratio ${\Sigma_c}/{\Sigma_h}$ is relatively large while it approaches the lower bound $\varepsilon_- \equiv 0$ when the ratio ${\Sigma_c}/{\Sigma_h}$ is small enough. In addition, the curve with parameter ${\Sigma_c}/{\Sigma_h}=1$ corresponds to $\varepsilon_{CA}=\sqrt{1+\varepsilon_c}-1$, which is also located in the region bounded between $\varepsilon_-$ and $\varepsilon_+$.

\begin{figure}[!htp]
\includegraphics[width=7cm]{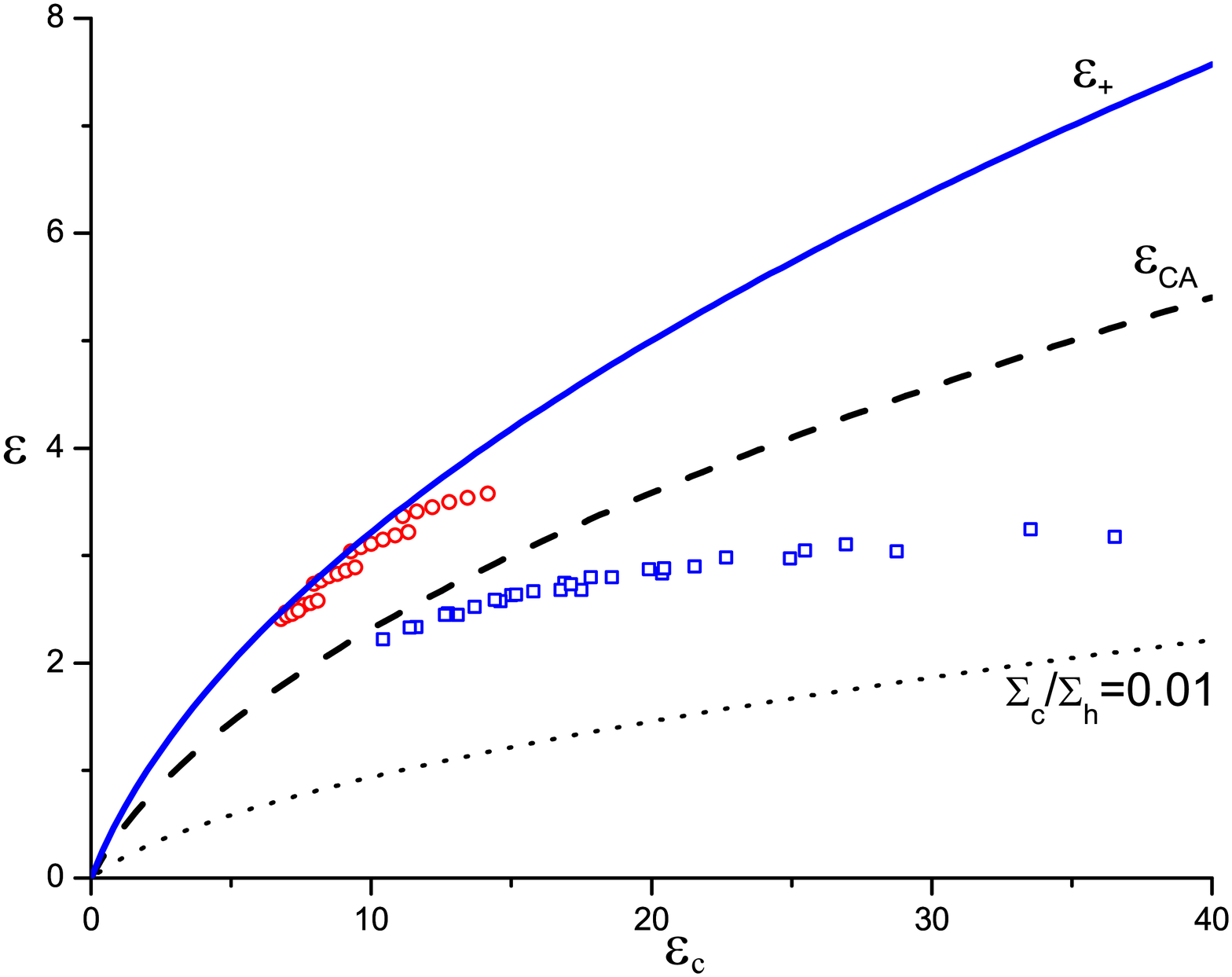}
\caption{(Color online) Comparison between theoretical prediction and the observed COPs of refrigerators.
The circles represent the relationship between observed COPs and the Carnot COPs calculated according to the working temperature region for the reciprocating chiller with nominal cooling rate 1172~kw while the squares represent that for  the water cooled reciprocating chiller with nominal cooling rate 10.5~kw~\cite{Gordon}. The solid line represents the theoretical upper bound $\varepsilon_+=(\sqrt{9+8\varepsilon_c}-3)/2$.}\label{graph2}
\end{figure}

Now we compare our prediction with the observed COPs of some real refrigerators. The circles and squares in Fig.~\ref{graph2} respectively show the relationship between the observed COPs of two different kinds of real refrigerators working in different temperature regions and the corresponding Carnot COPs calculated according to the working temperatures. The raw data are adapted from Tables 6.1 and 10.1 in Ref.~\cite{Gordon}. We stress from this figure that all data are located between the optimized COPs at $\Sigma_c/\Sigma_h\rightarrow \infty$ (the solid line) and
$\Sigma_c/\Sigma_h=0.01$ (the dotted line), which reveals the capability of the
low-dissipation assumption and the bounds of the optimized COP in
order to reasonably estimate the experimental results for real
refrigerators. Additionally, we also plot $\varepsilon_{CA}=\sqrt{1+\varepsilon_c}-1$ as the dashed line in Fig.~\ref{graph2}, from which we see that $\varepsilon_{CA}$ is neither the upper bound nor lower bound of observed COPs.
This result suggests that $\chi=\varepsilon Q_c/t_{\mathrm{cycle}}$ is indeed a very valuable figure of merit in comparing with experimental refrigerators data.

\emph{Conclusion}.-- The issue of COP at maximum figure of merit for Carnot-like refrigerators is addressed. We obtain the universal lower and upper bounds of COP at maximum figure of merit for low-dissipation Carnot-like refrigerators.
These bounds can be reached for extremely asymmetric dissipation cases. We compare our prediction with the observed COPs of real refrigerators and find that all measured COPs are located in between the prediction model. From a theoretical point of view, these results for low-dissipation refrigerators can be regarded as a counterpart of the bounds of efficiency at maximum power output obtained by Esposito \emph{et al.}~\cite{Esposito2010} for low-dissipation heat engines. In the future work, we will extend our discussions to the refrigerators working out of the low-dissipation regime based on the key equation~\eqref{varepsilonm} and our previous investigation on heat engines~\cite{wangtu12arxiv}.

\emph{Acknowledgement}.--The authors are grateful for the financial supports from the
National Natural Science Foundation of China (Grant No. 11075015), the Ministerio de Educacion y Ciencia of Spain (Grant FIS2010-17147-Feder)
and the Fundamental Research Funds for the Central Universities.
ZCT is also grateful for Yann Apertet and Jianhui Wang for their instructive discussions.

\end{document}